\documentclass[aps,prl,showpacs,twocolumn,superscriptaddress,
nofootinbib]{revtex4}
\usepackage[dvips]{epsfig}
\usepackage{graphicx}
\usepackage{amsmath}    
\newcommand{\ec}[1]{Eq.~(\ref{eq:#1})}
\newcommand{\eql}[1]{\label{eq:#1}}

\begin{document}


\title{Neutrinoless Universe}

\author{John F. Beacom}
\affiliation{NASA/Fermilab Astrophysics Center, Fermi National
Accelerator Laboratory, Batavia, Illinois 60510-0500}

\author{Nicole F. Bell}
\affiliation{NASA/Fermilab Astrophysics Center, Fermi National
Accelerator Laboratory, Batavia, Illinois 60510-0500}

\author{Scott Dodelson}
\affiliation{NASA/Fermilab Astrophysics Center, Fermi National
Accelerator Laboratory, Batavia, Illinois 60510-0500}
\affiliation{Dept. of Astronomy and Astrophysics,
The University of Chicago, Chicago, IL 60637}

\date{29 April 2004}

\begin{abstract}
We consider the consequences for the relic neutrino abundance if extra
neutrino interactions are allowed, e.g., the coupling of neutrinos to
a light (compared to $m_\nu$) boson.  For a wide range of couplings
not excluded by other considerations, the relic neutrinos would
annihilate to bosons at late times, and thus make a negligible
contribution to the matter density today. This mechanism evades the
neutrino mass limits arising from large scale structure.
\end{abstract}

\pacs{95.35.+d, 98.80.-k, 14.60.Pq, 13.35.Hb
\hspace{3cm} FERMILAB-Pub-04/050-A}


\maketitle


{\bf Introduction.---}
The discovery of neutrino oscillations means that neutrinos have mass,
which requires physics beyond the Standard Model.  The solar and
atmospheric oscillation experiments have measured neutrino
mass-squared differences $\delta m_{21}^2 \simeq 7 \times 10^{-5}
{\rm\ eV}^2$ and $\delta m_{32}^2 \simeq 2 \times 10^{-3} {\rm\
eV}^2$~\cite{OscResults}, which implies {\it lower} limits on 
two neutrino masses of $\sqrt{\delta m^2_{21}}$ and $\sqrt{\delta
m^2_{32}}$.  Since these oscillations have been shown to be dominated
by active-flavor neutrino oscillations, the three neutrino masses are
connected and become degenerate in mass if any are larger than
$\sqrt{\delta m^2_{32}}$~\cite{SolarDecay}.  Thus, at the present
sensitivity of $m_\nu < 2.2$ eV (at 95\% CL)~\cite{Tritium}, the {\it
upper} limit on neutrino mass from tritium beta decay applies to {\it
each} of the three mass eigenstates.  KATRIN, a proposed
next-generation tritium beta decay experiment, will have sensitivity
down to $m_\nu \simeq 0.2$ eV~\cite{Katrin}.  New neutrinoless double
beta decay experiments will have even greater sensitivity, but only if
neutrinos are Majorana particles~\cite{ElliottVogel}.

Neutrino mass can also be measured with cosmological data.  When
neutrinos are relativistic, they free stream out of density
perturbations, reducing the growth of structure.  This results in a
suppression of the matter power spectrum on all scales below that of
the horizon at the time the neutrinos became non-relativistic, after
which they act like cold dark matter.  The extent to which this lack
of clustering affects the distribution of matter today depends on the
ratio of the energy density of the non-clustering component
(neutrinos) to the total density of matter. The former is
\begin{equation}
\label{eq:omega}
\rho_\nu 
= \Sigma m_\nu n_\nu
= \frac{\Sigma m_\nu}{93.5 h^2 {\rm\ eV}}\,\rho_{\rm cr}\,,
\end{equation}
where $\rho_{\rm cr}=3H_0^2/8\pi G$ is the critical density associated
with a flat universe; the total density in matter is parametrized as
$\Omega_m \rho_{\rm cr}$. Here $h$ specifies the Hubble constant,
$H_0=100 h\, {\rm km\,sec}^{-1} \,{\rm Mpc}^{-1}$.  The equality on
the right in \ec{omega} assumes the standard cosmological
abundance. Recall that, in the standard scenario, neutrinos couple to
the rest of the cosmic plasma until the weak interactions freeze out
at $T\sim 1$ MeV. After neutrinos freeze out, their abundance scales
simply as $a^{-3}$ where $a$ is the cosmic scale factor. Thus, in the
standard cosmology, there are roughly as many relic neutrinos today as
photons in the cosmic microwave background (CMB).

Limits from structure formation on the sum of neutrino masses now
range from $0.5$ to $2$ eV, with the spread largely due to different
assumptions about the relative bias between the mass and galaxy
distributions~\cite{CosmoMass}.  Bias is one important issue, but this
will be circumvented with future weak lensing surveys, which will
measure the mass distribution directly.  Indeed, it has been shown
that these observations should realistically be able to reach the
scale $\sqrt{\delta m^2_{23}}$, by which the discovery of neutrino
mass is guaranteed~\cite{Guaranteed}.  These mass constraints depend
on assuming the standard relic neutrino abundance.  Big-bang
nucleosynthesis (BBN) constraints, combined with neutrino mixing data,
no longer allow the possibility of a significantly increased $n_\nu$
due to a large lepton asymmetry~\cite{Degen}.  Are there other ways to
alter the relic neutrino abundance, and specifically to lower it?

If neutrinos have extra interactions so that they remain in
equilibrium until late times, they would freeze out when they are
non-relativistic, in which case their final abundance would be
suppressed by a factor $\propto e^{-m_\nu/T_f}$.  We show that new
neutrino couplings in the allowed range can lead to a {\it vanishing
relic neutrino density today}, hiding the effects of neutrino masses
from cosmological observations.  This possibility is falsifiable both
directly and with other experiments.


{\bf Interaction model.---}
We consider the cosmological consequences of coupling neutrinos to
each other with bosons, through tree level scalar or pseudoscalar
couplings of the form
\begin{equation}
\label{coupl}
{\cal L}= h_{ij}\overline{\nu}_i \nu_j \phi + 
g_{ij} \overline{\nu}_i \gamma_5 \nu_j \phi + {\rm h.c.},
\end{equation}
as in Majoron-like models, for example.  The field $\phi$ is assumed
to be massless (or light compared to $m_\nu$).  Viable models of this
type have been discussed in Ref.~\cite{Models}.  Here we assume that
there is just one new boson, and that these new couplings are
unconnected to the mechanism of neutrino mass generation.  Even tiny
couplings can cause profound effects, as we will show.

The solar neutrino~\cite{SolarDecay,SolarDecay2} and meson
decay~\cite{MesonDecay} limits on these couplings are very weak, $|g|
\lesssim 10^{-2}$ (here and below we do not distinguish $g$ or $h$
type couplings, nor neutrinos and antineutrinos).  Neutrinoless double
beta decay limits $g_{ee} < 10^{-4}$, but the other couplings may be
much larger.  Supernova constraints may exclude a narrow range of
couplings around $g \sim 10^{-5}$, but the boundaries are model
dependent~\cite{Supernova}.  Scalar couplings could mediate long-range
forces with possible cosmological consequences, while pseudoscalar
couplings mediate spin-dependent long-range forces, which have no net
effect on an unpolarized medium~\cite{LongRange}.  Since these
constraints can be evaded, and since in our case $\phi$ couples only
to neutrinos, we do not consider them further.

The $\phi$ boson can be brought into thermal equilibrium through its
coupling to the neutrinos, and the $\nu - \phi$ system may stay in
thermal contact until late times, through the processes $\nu \phi
\leftrightarrow \nu \phi$ and $\nu \leftrightarrow \nu \phi$.  Most
important though is $\nu \nu \leftrightarrow \phi \phi$, a process
which depletes the total number of neutrinos.  In the standard case,
the neutrinos decouple from each other and the matter at $T \sim 1$
MeV, but interactions with $\phi$ may keep neutrinos in equilibrium
until they are non-relativistic, $T \sim 1$ eV, when the inverse
process becomes kinematically prohibited.  In order to accomplish this, $g$
must be sufficiently large; we show below that this requires $g
\gtrsim 10^{-5}$, well within the allowed range.  While neutrino decay
requires off-diagonal couplings, the effects considered here can occur
with either diagonal or off-diagonal couplings.  If the couplings are
this large, all relic neutrinos efficiently annihilate into
bosons, leaving no relic neutrinos today, thereby hiding the cosmological 
effects of neutrino mass.

Past models of invisible neutrino decay also allowed a late transfer
of energy from non-relativistic to relativistic particles, altering
the expansion rate history~\cite{CosmoDecay}.  However, the case
considered was that of a heavy ($m_\nu \gtrsim 10$ eV) neutrino,
enough to be the dark matter, decaying into massless neutrinos.  Such
scenarios are no longer possible, given laboratory data on the
neutrino mass scale and mass differences.  For the relevant mass
range, if decays occur, the parent and daughter neutrinos are equally
relativistic.  The possibilities of neutrino annihilation and/or
self-interaction have been considered in scenarios in which neutrinos
are the dark matter~\cite{CosmoAnnil,Kolb}.  When neutrinos are a
fraction of the dark matter, the
signatures of this scenario are more subtle, and have not been treated
elsewhere.


{\bf Annihilation.---}
The neutrino annihilation rate is
\begin{equation}
\Gamma = \langle \sigma v \rangle n_{\rm eq},
\end{equation}
with the cross section~\cite{Crosssection,Kolb}
\begin{equation}
\label{sigma}
\sigma = \frac{g^4}{32 \pi} \frac{1}{s} \left[ 
\frac{1}{\beta^2} \log\frac{1+\beta}{1-\beta} -\frac{2}{\beta} \right],
\end{equation}
where $\sqrt{s}$ is the center of mass energy and $\beta^2 =
1-4m_\nu^2/s$.  In the non-relativistic limit the annihilation rate
becomes
\begin{equation} 
\Gamma(T) = \frac{g^4 }{64 \pi } \frac{T}{m_\nu^3} 
\left( \frac{m_\nu T}{2\pi} \right)^{3/2} e^{-m_\nu/T}, 
\eql{annrat}\end{equation}
where we have used $\langle\beta^2\rangle \simeq 3 T/m_\nu$.

For sufficiently large $g$, the annihilation rate will be larger than
the expansion rate until the temperature drops well below the
neutrino mass.  Once $T_\nu < m_\nu $, the neutrino abundance will become
exponentially suppressed, asymptoting to the equilibrium abundance at
the freezeout temperature, $T_f$, defined as the temperature at which
the annihilation rate is equal to the expansion rate.  If $T_f$ is
less than of order $m_\nu/7$, the neutrinos will be suppressed from their
nominal abundance by a factor greater than $100$: they will play no
role in subsequent cosmological evolution.  The constraint on the
coupling $g$ is thus obtained by solving $\Gamma(T_f) \equiv H(T_f)$
and requiring $T_f<m_\nu/7$.  For this estimate, it is sufficient to set
$H(T) \sim H_0 [\Omega_m (T/T_0)^3 + \Omega_\gamma (T/T_0)^4]^{1/2}$,
where $T_0$ is the standard photon temperature, $2.73$K, and
$\Omega_\gamma = 2.47\times 10^{-5} h^{-2}$ is the ratio of energy
density in photons to the critical density~\cite{Scott}.  Then, we
find that as long as $g \agt 10^{-5}$, the annihilation is complete by
$T_f$, with only a negligible amount of neutrinos remaining.

Note that for $g \agt 10^{-5}$, the boson will be brought into thermal
equilibrium before BBN.  The energy density of a scalar boson is equivalent
to 4/7 that of a neutrino species.  Current BBN
limits~\cite{Olive,Marfatia,Cuoco} are $N_\nu^{\rm eff} < 3.3 - 4$, so
an additional boson is still allowable.  In the case that the electron
neutrinos have a large lepton asymmetry, even $N_\nu^{\rm eff}=7$ is
permitted, provided the extra degrees of freedom do not consist of
active neutrinos~\cite{Degen,Marfatia}.  Neutrino-majoron interactions
may also weaken the constraints on large lepton
asymmetries~\cite{Dolgov}.


{\bf Neutrino-boson energy density.---}
We shall henceforth assume that $g>10^{-5}$, so the neutrinos completely 
annihilate into massless bosons.  However, there is 
still a small impact on the distribution of matter in the
universe today.
%
%
The energy density in the $\nu-\phi$ system differs
from that of the three massless neutrinos of the canonical standard
cosmological model and from a model of three massive non-interacting
neutrinos. In particular, the epoch of matter domination is delayed in
the interacting neutrino scenario outlined above. This delay leads to
a small suppression of the matter power spectrum on small scales. To
explain this suppression, we first compute the evolution of the energy
density in the $\nu-\phi$ system and compare it with the conventional
scenarios.

\begin{figure}
\includegraphics[width=3.25in]{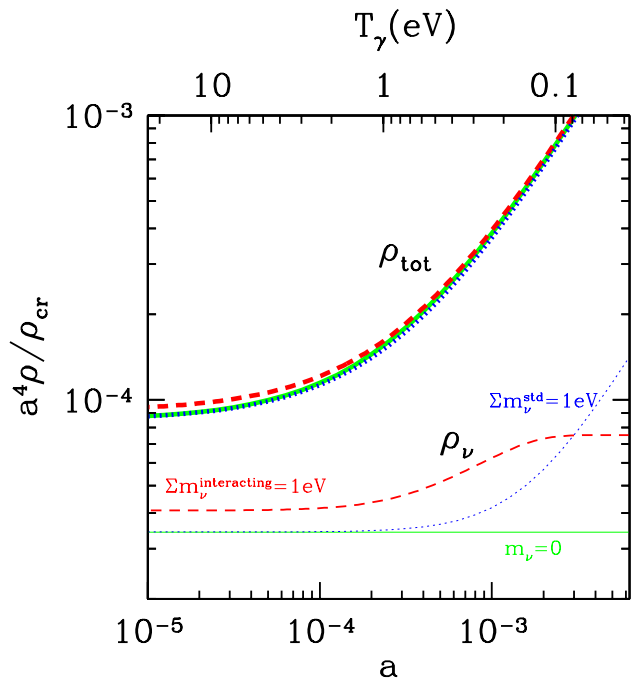}
\caption{\label{fig:energy} Evolution of the energy density 
as a function of the scale factor $a$.  Heavy curves at top are total 
energy density including matter, photons, and neutrinos; 
light curves at bottom are energy density in the neutrino
sector (including  $\phi$'s in the interacting case). Three
different scenarios are depicted, differing in neutrino content: three
massless neutrinos (solid), three degenerate standard model
neutrinos with $\sum m_\nu = 1$ eV (dotted); and
three interacting degenerate neutrinos plus massless $\phi$
(dashed). 
We use the same total matter density, $\Omega_m =0.3$, throughout;
$\rho_{\text cr}$ denotes the critial debsity {\it today}. 
}
\end{figure}

As the neutrinos annihilate, the common temperature of the $\nu-\phi$
fluid does not simply scale as $a^{-1}$.  Rather, it falls less
sharply.  To track the temperature evolution, we can use entropy
conservation.  The entropy density of the $\nu-\phi$ fluid is
\begin{equation}
s_{\nu-\phi} = \frac{2 \pi^2}{45}T_{\nu\phi}^3 \left[1+6 \times(7/8) 
F(m_\nu/T_{\nu\phi})\right],
\end{equation}
where
\begin{equation}
F(m_\nu/T_{\nu\phi}) \equiv \frac{180}{7 \pi^2 T_{\nu\phi}^4} 
(\rho_{\nu} + P_{\nu})
.\end{equation} When the neutrinos are highly relativistic, $F=1$,
while it is exponentially suppressed, $F \simeq 0$,  at late times when 
the neutrinos become non-relativistic.
Entropy conservation then implies
\begin{equation}
\label{Tnu}
\frac{T_{\nu\phi}}{T_\gamma}
= \left( \frac{T_{\nu\phi}}{T_\gamma} \right)_{\rm init}
\left[ \frac{1+21/4}{1+(21/4)F(m_\nu/T_{\nu\phi})} \right]^{1/3}.
\end{equation}
If $(T_{\nu\phi}/T_\gamma)_{\rm init}$ takes the standard value,
$(4/11)^{1/3}$ at early times, at late times we have
$(T_{\nu\phi}/T_\gamma)=(25/11)^{1/3}$.  This implies an increase in
the radiation energy density, corresponding to an effective number of
neutrinos of $N_\nu^{\rm eff} = 6.6$.  The evolution of the energy
density is shown in Fig.~\ref{fig:energy}.


CMB measurements constrain the number of light relativistic degrees of
freedom. The current limit is $N_\nu^{\rm eff} \alt 7$~\cite{Crotty},
hence does not rule out this scenario. 
Further, one must be careful about applying this limit to our model,
as interactions will reduce the propagation 
speed of neutrinos. 
Some secondary effects on the CMB due to neutrino freestreaming 
(i.e., a phase shift and amplitude reduction) will thus
be less striking than in the standard, non-interacting model~\cite{Seljak}. 
These effects are discussed in \cite{Chacko} for a 
similar model in which a light (but heavier than $m_\nu$) boson is coupled 
to the neutrinos.


\begin{figure}
\includegraphics[width=3.25in]{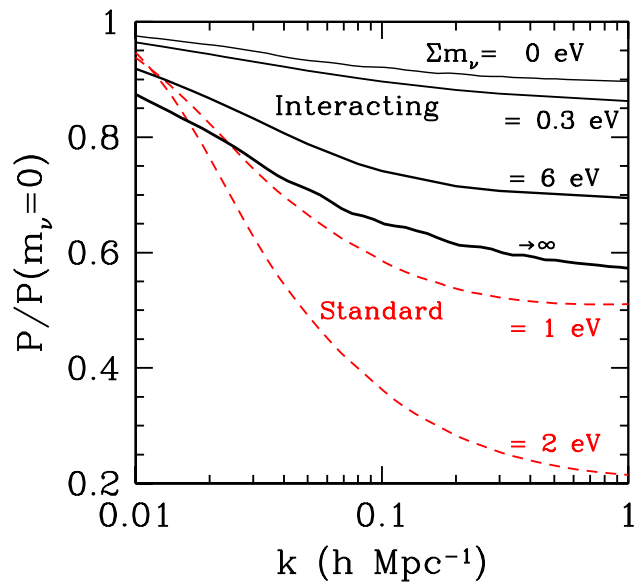}
\caption{\label{fig:diff} The ratio of power spectra $P/P(m_\nu=0)$
where $P(m_\nu=0)$ is the power spectrum for the standard scenario
with massless neutrinos.  The solid curves show this ratio for various
(degenerate) neutrino masses in the interacting scenario. Dashed curves show
the ratio in the standard scenario, for which the   
current limit is $\sum m_\nu < 1-2$ eV.
Note that the tritium bound, $\sum m_\nu < $ 6.6 eV, always applies. 
}
\end{figure}

{\bf Power spectrum.---}
We have calculated the large scale structure power spectrum, assuming
the limit where the neutrino annihilation is complete.  We find that
the current neutrino mass limits can be completely removed: all values
of $\sum m_\nu$ are allowed, even those much greater than 1 eV.  The
result are shown in Fig.~\ref{fig:diff} where for comparison we have
also shown the suppression caused by free streaming in the standard
case.

In the interacting scenario, the usual suppression due to neutrino
mass is absent, because neutrinos make no contribution to the
matter density today.  A small suppression does
occur, due to the extra radiation present.  Even though neutrinos do
not freestream, perturbations in the neutrino-$\phi$
fluid still cannot grow, due to the pressure in this tightly coupled
relativistic fluid.  The negligible density in neutrinos makes this
suppression irrelevant.  The effects on the power spectrum are thus
entirely due to the modified expansion history.

Matter radiation equality is delayed, since the $\phi$ heating leads to an
enhanced radiation density (see Fig.~1).
Therefore, the
potentials for scales which enter the horizon during the radiation
dominated epoch will decay for a slightly longer period, leading to a
small suppression of the power spectrum on these scales.  Note that if
the neutrino annihilation is complete well before matter radiation
equality, as would be the case for very heavy neutrinos, the full
effects of the extra radiation are felt.  This corresponds to the
bottommost of the solid curves in Fig.\ref{fig:diff}.  For very small
neutrino masses, $m_\nu \ll 1 {\rm eV}$, the increase in the radiation
density due to neutrino annihilation occurs after time of matter
radiation equality.  At this stage, the universe has already entered
the matter dominated regime, where the potentials are dominated by the
dark matter, and the radiation is less important.  The effects of the
extra radiation created by neutrino annihilation are thus quite
small. (The power spectrum is slightly suppressed with respect to a
standard massless neutrino scenario, since there is still a small
amount of extra radiation due the population of $\phi$.) For
intermediate cases, e.g., $\sum m_\nu = 1 {\rm eV}$, we find a
suppression $P/P(m_\nu=0) \simeq 0.8$, compared to 0.5 in the normal
case.


{\bf Conclusions.---}
We have examined a model in which extra couplings allow the neutrinos
to annihilate into massless (or light) bosons at late times, and thus
make a negligible contribution to the matter density today.  This
evades the present neutrino mass limits arising from large scale
structure.  Future tritium beta decay experiments like
KATRIN~\cite{Katrin} will play a unique and essential role, especially
in comparison to cosmology and neutrinoless double beta decay,
allowing stringent tests of new neutrino interactions.

The scenario outlined here could be falsified in several ways.  First,
by a robust discovery of neutrino mass with large scale structure
data, if the power spectrum suppression was greater than that allowed
for the tritium bound mass in the interacting case (see
Fig.~\ref{fig:diff}).  This emphasizes the importance of improving the
tritium bound.  Second, with future precision BBN and CMB data.
Third, these couplings could lead to neutrino decay over astronomical
distances, which has testable consequences~\cite{Astrodecay}.


%
We thank Kev Abazajian, Marc Kamionkowski, Andy Rawlinson, and Mike
Turner for discussions, and acknowledge support by Fermilab (DOE
Contract No. DE-AC02-76CH03000) and NASA Grant No. NAG5-10842.


\end{document}